\documentclass[10pt, conference, compsocconf]{IEEEtran}

\usepackage{amsmath}
\usepackage{graphicx}
\usepackage{courier}
\usepackage{float}
\usepackage{subfig}
\usepackage{flexisym}
\usepackage{tabularx}
\usepackage{pdfcomment}
\usepackage{algpseudocode}
\usepackage{algorithm}

\DeclareMathOperator*{\argmin}{arg\,min}
\DeclareMathOperator*{\argmax}{arg\,max}
\algnewcommand{\algorithmicgoto}{\textbf{go to}}%
\algnewcommand{\Goto}[1]{\algorithmicgoto~\ref{#1}}%

\begin{document}

\title{Executing Bag of Distributed Tasks on the Cloud:\\Investigating the Trade-offs Between Performance and Cost}

\author{\IEEEauthorblockN{Long Thai, Blesson Varghese and Adam Barker}
\IEEEauthorblockA{School of Computer Science, University of St Andrews, Fife, UK\\
Email: \{ltt2, varghese, adam.barker \}@st-andrews.ac.uk}
}

\maketitle

\begin{abstract}
Bag of Distributed Tasks (BoDT) can benefit from decentralised execution on the Cloud. However, there is a trade-off between the performance that can be achieved by employing a large number of Cloud VMs for the tasks and the monetary constraints that are often placed by a user. The research reported in this paper is motivated towards investigating this trade-off so that an optimal plan for deploying BoDT applications on the cloud can be generated. A heuristic algorithm, which considers the user's preference of performance and cost is proposed and implemented. The feasibility of the algorithm is demonstrated by generating execution plans for a sample application. The key result is that the algorithm generates optimal execution plans for the application over 91\% of the time. 
\end{abstract}

\IEEEpeerreviewmaketitle

\section{Introduction}
\label{introduction}

\textbf{Bag of Tasks} (BoT) refers to a collection of independent and identical tasks, which can be executed in sequence or in parallel. 
There is a subset of BoT, named \textbf{Bag of Distributed Tasks} (BoDT), in which each task additionally requires data placed at different geographical locations for execution.

\begin{figure}[h!]
\subfloat[BoT]     {\includegraphics[width = 0.24\textwidth]{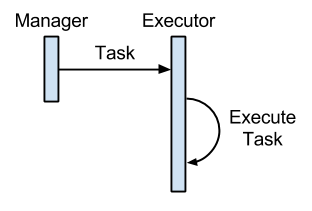}\label{bot}} \hfill
\subfloat[BoDT]    {\includegraphics[width = 0.24\textwidth]{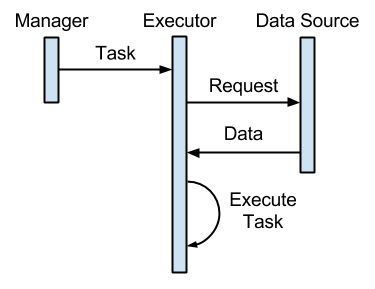}\label{bodt}} \hfill
\caption{Difference between BoT and BoDT in executing a task}
\label{bot-bodt}
\end{figure}

A manager and an executor are required by both BoT and BoDT for executing tasks. The key difference is illustrated in Figure \ref{bot-bodt}. In BoT, a task is executed as soon as it is received by an executor. In BoDT, an executor has to request the data required by the task from a globally located source before it can be executed. This step makes scheduling the execution more complicated as the location of data sources must be included into the scheduling algorithm, which incurs an additional cost since the time required for transferring data utilised by the task has to be taken into account in addition to the time taken for processing the task.


Centralised approaches suffer high communication costs between the executor and the data sources as all tasks are executed at one fixed location. Since a BoDT consists of independent tasks, it is possible to divide it into smaller sub-sets based on the tasks' geographical distribution and execute each of them on an application located near their data sources. This is a decentralised approach, which can potentially reduce the communication cost.

Cloud computing offers the infrastructure to decentralise the execution of BoDT as its providers maintain multiple data centres, which are geographically distributed across the world. It is possible to rapidly deploy an application on a data centre close (in terms of network distance) to its data sources in order to reduce data transfer times. 

Consider for example BoDT applications, such as Feedly\footnote{http://feedly.com/} and Flipboard\footnote{https://flipboard.com/}, that aggregates news items for a user by collecting them from globally distributed sources. The application can benefit from being executed on multiple Cloud data centres in a manner that will reduce the network distance between it and the data sources.

While data transfer times can be reduced in BoDT on the cloud, a different kind of a problem emerges since a user pays for the cloud resources required by the tasks. Often more resources will boost the performance of the task, but this comes at a monetary cost. Hence, there is a trade-off between the performance gain by acquiring more resources on the cloud for executing a task and the budget available for executing the task on the cloud. For example, the performance of the BoDT applications that aggregate news considered above can be maximised by gathering more resources for increasing parallelism. However, this is more expensive and may not be cost-effective.   

In this paper, we set out to investigate the trade-off between performance gain and monetary cost of executing BoDT on the Cloud. To facilitate the investigation a heuristic algorithm is proposed and developed, which takes into account a user's preference of performance over cost specific to an application and/or the user's desired Service Level Agreement (SLA). The algorithm generates a execution plan for BoDT applications based on the user preference. The feasibility of the algorithm is demonstrated using a sample user application and the key result is that for 10 out of 11 times an accurate execution plan is generated. 

The contributions of the research reported in this paper are (i) providing a mathematical model of executing BoDT on the Cloud, (ii) introducing a heuristic algorithm to find the execution plan based on a user's preference, and (iii) investigating the trade-off between performance and cost in executing BoDT on the Cloud.

The remainder of this paper is organised as follows. Section \ref{problemmodelling} introduces the mathematical models of executing BoDT. Section \ref{heuristicalgorithms}, presents a heuristic algorithm for generating an execution plan. Section \ref{experimentalstudies} shows the experimental evaluation of the framework. Section \ref{relatedwork} highlights the related work. Section \ref{conclusions} concludes this paper.

\section{Problem Modelling}
\label{problemmodelling}

In this section, we present mathematical models for centralised and decentralised execution of BoDT, and a discussion on the trade-off between performance and cost for the decentralised execution model. 

\subsection{Centralised Execution}

In the centralised execution model of BoDT on the cloud only one VM is used. The cost of this VM depends on its running time. Hence, an optimal execution model is obtained when a location to deploy a BoDT application is found such that the running time is at its minimum.

Let $T = \{t_1, t_2... \}$ be the collection of tasks and $L = \{l_1, l_2...\}$ be the locations of the tasks. Each task can be represented as a pair that describes it location and the size of the data, i.e. $t = (l_t, s_t)$ for $t \in T$, $l_t \in L$ and $s_t > 0$.

Let $C = \{c_1, c_2...\}$ be the collection of Cloud VMs for deploying the BoDT. For simplicity, we assume that each region can only host the same type of VM. Deploying different VM types simultaneously in a region will be investigated in the future and reported elsewhere.

For the task location $l \in L$ and the VM $c \in C$, the time taken for transferring one unit of data is denoted as $trans_{l, c}$. Hence, the time taken to transfer the data of a task $t \in T$ to the VM $c \in C$ is:

\begin{equation*}
	trans_{t, c} = s_t \times trans_{l_t, c}
\end{equation*}

The performance of each VM $c \in C$ in the model is indicative of how long it takes to process one unit of data and is represented as $comp_c$. Since this model supports VMs of the same type in a region the performance of the VMs is identical and is denoted as $comp$. The time taken to execute a task $t \in T$ on the VM $c \in C$ is:

\begin{equation*}
	comp_{t, c} = s_t \times comp
\end{equation*}

The execution time of task $t$ on VM $c$ is:

\begin{equation*}
	exec_{t, c} = trans_{t, c} + comp_{t, c}
\end{equation*}

The total time taken to execute all tasks $t \in T$ on VM $c$ is:

\begin{equation*}
	exec_{T, c} = \sum_{t \in T}{exec_{t, c}}
\end{equation*}

In order to maximise the performance, the VM should be selected so that the total execution time $exec_{T, c}$ is minimum. As the computational performance between VMs is identical, the total execution time is minimised when the data transferring time is minimised. The optimal Cloud VM $c_o$ is selection as:

\begin{equation*}
	c_o = \argmin_{c \in C}{(exec_{T, c})}
\end{equation*}

The running time of a Cloud VM is the sum of the total execution time and the deployment time $dt$, which is the time taken to start up and deploy the application on a VM. Notably, the Cloud VM is only started if and only if there is tasks assigned to it, i.e. its execution time is larger than 0. Otherwise, the Cloud VM will not be started, which means its running time is 0. Hence, the running time is calculated as:

\begin{equation} \label{running_time}
	running\_time_{T,c} =\left\{
		\begin{array}{l l}
		dt + exec_{T,c} & \quad \text{if $exec_{T,c} > 0$}\\
			0 & \quad \text{otherwise}
		\end{array} \right.
\end{equation}

Cloud VMs are charged for the time blocks used and is usually by the hour (equivalent to 3600 seconds). A time block running the BoDT $T$ on VM $c$ is represented as:

\begin{equation} \label{time_block}
	tb_{T,c} = \lceil \frac{running\_time_{T,c}}{3600} \rceil
\end{equation}

The cost of each time block is identical and denoted by $unit\_cost$. Therefore the total cost is 

\begin{equation} \label{instance_cost}
	cost_{T,c} = tb_{T,c} \times unit\_cost
\end{equation}

This cost can be minimised by selecting a Cloud VM which takes the lowest time for executing the task. 

\subsection{Decentralised Execution}

In the decentralised execution model the BoDT is divided into groups of tasks, each of which is executed by one VM.

For $c \in C$, $T_c \subset T$ denotes all tasks executed on $c$, we have two constraints. Firstly, the constraint that ensures that all tasks are executed, which is:

\begin{equation} \label{completeness}
	\cup_{c \in C}{T_c} = T
\end{equation}

Secondly, the constraint that ensures that each task is executed only on one VM, which is:

\begin{equation} \label{unity}
	T_i \cap T_j = \emptyset \quad \text{for $i, j \in C$ and $i \neq j$}
\end{equation}





The execution time for each VM is:

\begin{equation*}
	exec_{c} = exec_{T_c} = \sum_{t \in T_c}{exec_{t, c}}
\end{equation*}

The $running\_time$ of each VM is calculated similar to Equation \ref{running_time}. Hence, the total number of used time blocks is:

\begin{equation} \label{total_cost}
	tb_{T, C} = \sum_{c \in C}{\lceil \frac{running\_time_{c}}{3600} \rceil}
\end{equation}

Since all VMs run in parallel, the overall time taken for executing all tasks is:

\begin{equation} \label{overall_exec}
	overall\_exec = \max_{c \in C}{exec_c}
\end{equation}

As stated in Equation \ref{overall_exec}, the $overall\_exec$ depends on the VM with \textbf{highest execution time}. In other words, it does not matter if one VM manages to finish its execution very quickly, as long as the other are still running. Hence, it is beneficial to balance the tasks between VMs so that they can finish approximately at the same time.

\subsection{Discussion} \label{sec:model_discussion}

Performance of the execution of BoDT on the Cloud can be maximised by minimising the overall execution time presented in Equation \ref{overall_exec}. This can be achieved by using multiple Cloud VMs so that many tasks can be executed in parallel and the load of task execution on each VM is reduced. Cloud VMs are charged by the hour and therefore using a large number of VMs may not be cost effective if they are all required only for a small fraction of the hour. Hence, tasks must be scheduled such that the total number of charged time blocks as shown in Equation \ref{total_cost} are minimised. Moreover, data transfer times can be reduced by assigning task to Cloud VMs closest to the data source. However, data can be distributed all over the world which may result in a VM being assigned more tasks than others. This problem can be circumvented with the help of a balancing mechanism which will be presented in the next section.

There is a trade-off in minimising the total cost and the overall execution times shown in Equation \ref{total_cost} and Equation \ref{overall_exec}. A user must decide whether the BoDT application is executed for maximum performance, or for minimum costs, for optimality by taking both reasonable performance and cost effectiveness into account. If $\beta$ is a factor that represents the importance of minimising overall execution time, and $0 \leq \beta \leq 1$, then the optimal plan satisfies the unity constraints shown in Equations \ref{completeness}, \ref{unity} and minimises: 

\begin{equation} \label{score}
	score = \beta \times \overline{overall\_exec} + (1 - \beta) \times \overline{tb_{T, C}}
\end{equation}

The overall execution times and number of time blocks of each instance are normalised so that both of them have the same scale. 

\section{Heuristic Algorithms}
\label{heuristicalgorithms}

Minimising Equation \ref{score} while satisfying all the constraints is a constraint programming problem. Although solving this problem is able to produce an optimal solution, it is expensive and can take a considerable amount of time until a final answer is found. Hence, in this section, a heuristic algorithm is proposed in order to find a plan based on given $\beta$ value in a more efficient way.

\subsection{Nearest Plan}

Due to the homogeneity of cloud resources, the time to process the same amount of data between all locations should be identical. The difference in execution time comes from transferring data from its source to a VM. Intuitively, each task should be executed at the location nearest to its data source in order to minimise the communication time, thus minimising the total execution time. The plan in which all tasks are assigned to locations nearest to their data sources is \textbf{nearest plan} and presented in Algorithm \ref{nearest_plan}.

\begin{algorithm}
	\caption{Get Nearest Plan}
	\label{nearest_plan}
	\begin{algorithmic}[1]
		\Function{GET\_NEAREST\_PLAN}{$T, C, trans$}
			\State // create an empty list of tasks for each cloud VMs
			\For{$c \in C$}
				\State $T_c \gets \emptyset$
			\EndFor
			\For{$t \in T$}
				\State $c = \argmin_{c \in C}{trans_{t, c}}$ // find the nearest VM
				\State $T_c \gets T_c \cup \{t\}$
			\EndFor
			\State $T_n \gets \{T_c$ for $c \in C$\}
			\State \Return $T_n$
		\EndFunction
	\end{algorithmic}
\end{algorithm}

\subsection{Cost Effective Plan}

\begin{algorithm}
	\caption{Reduce Time Block}
	\label{reduce_time_blocks}
	\begin{algorithmic}[1]
		\Function{REDUCE\_TB}{$trans, T_n, min\_tb$}\label{reduce:start}
			\State $T_{ce} \gets T_n$
			\State $T_{c_0} \gets \argmin_{T_c \in T_{ce}}{(running\_time_{T_c} \mod 3600)}$\label{reduce:select_vm}
			\State $t_0 \gets \argmax_{t \in T_{c_0}}{exec_{t,c_0}}$\label{reduce:select_task}
			\State $C_1 \gets (C - \{c_0\})$ ordered by $trans_{t_0, c}$\label{reduce:sort_vms}
			\For{$c_1 \in C_1$}\label{reduce:loop}
				\If{$t_0$ is never moved from $c_0$ to $c_1$}\label{reduce:no_repeat}
					\State $tb\textprime_{c_1} := \lceil \frac{running\_time_{c_1} + exec_{t_0, c_1}}{3600} \rceil$\label{reduce:new_tb}
					\If{$tb\textprime_{c_1} = tb_{c_1}$}\label{reduce:compare_tb}
						\State $T\textprime_{c_0} \gets T_{c_0} - t_0$\label{reduce:mv_task_1}
						\State $T\textprime_{c_1} \gets T_{c_1} \cup \{t_0\}$\label{reduce:mv_task_2}
						\State $T_{ce} \gets (T_{ce} - \{T_{c_0}, T_{c_1}\}) \cup \{T\textprime_{c_0}, T\textprime_{c_1}\}$\label{reduce:mv_task_3}
						\If{$tb_{T} > min\_tb$}
							\State \Goto{reduce:select_vm}
						\EndIf
					\EndIf
				\EndIf
			\EndFor
			\State \Return $T_{ce}$
		\EndFunction
	\end{algorithmic}
\end{algorithm}

As mentioned in Section \ref{sec:model_discussion}, using as many VMs as possible is not cost effective, hence we develop an algorithm which reduces the number of charged time blocks of the $nearest\_plan$. Algorithm \ref{reduce_time_blocks} aims to find the \textbf{cost effective plan} which uses the minimum number of time blocks. The number of time blocks are minimised by filling each of them with as many tasks as possible.

Besides the distance between VMs and tasks, Algorithm \ref{reduce_time_blocks} takes as input the nearest plan produced by Algorithm \ref{nearest_plan} and the minimum number of time blocks allowed (Line \ref{reduce:start}). The most cost-ineffective VM whose execution time has the smallest fraction of an hour is selected (Line \ref{reduce:select_vm}). The selected VM has the highest charged but unused time.

The task with the highest execution time is selected from the chosen VM (Line \ref{reduce:select_task}). Next, the remaining cloud VMs are sorted in descending order based on their distance to the selected task's data (Line \ref{reduce:sort_vms}). Sorting is performed to assign tasks to the closest possible VM, thus minimising the execution time.

Then, the remaining sorted VMs are iterated through in order (Line \ref{reduce:loop}). A constraint in line \ref{reduce:no_repeat} is added in order to avoid an infinite loop in which a task is moved back and forth between 2 VMs.

As the algorithm aims to lower the number of time blocks by moving tasks from one VM to another, the number of time blocks used by the VM receiving a new task must not increase. In order to ensure this, the new number of time blocks if a task is assigned to the VM (Line \ref{reduce:new_tb}) is compared against the previous number of time blocks (Line \ref{reduce:compare_tb}).

If a VM satisfying all the constraints is found, the task is removed from the cost-ineffective VM (Line \ref{reduce:mv_task_1}) and added to the selected VM (Line \ref{reduce:mv_task_2}). Finally, the current plan, including the list of VMs and their assigned tasks, is updated (Line \ref{reduce:mv_task_3}).

The aim is to find a plan that reflects the trade-off between performance and cost instead of a plan that has the lowest number of time blocks. The value $min\_tb$ is used to limit the number of time blocks to accommodate the cost that a user is willing to pay. The number of time blocks of the nearest plan is the upper bound to $min\_tb$ with a lower bound of 1, assuming there are always tasks to be executed. 

By de-assigning a task from its current VM, the execution time is reduced, and consequentially its used time blocks. It is possible to remove all tasks from a certain VM, i.e the VM is not used any more. So the algorithm reduces not only the number of charged time blocks but also the number of VMs. This is beneficial as each VM has an initial deployment time that is charged but cannot be used for executing tasks. This means that if a user makes use of more VMs, there is likely to be more unused time.

\subsection{Balancing Algorithm}

\begin{algorithm}
	\caption{Balancing Algorithm}
	\label{balance}
	\begin{algorithmic}[1]
		\Function{BALANCE}{$trans, T_{ce}$}
			\State $T_{b} \gets T_{ce}$
			\State $T_{c_0} \gets \argmin_{T_c \in T_{b}}{exec_{T_c}}$\label{balance:select_vm}
			\State $t_0 \gets \argmax_{t \in T_{c_0}}{exec_{t,c_0}}$
			\State $C_1 \gets (C - \{c_0\})$ ordered by $trans_{t_0, c}$
			\For{$c_1 \in C_1$}
				\If{$t_0$ is never moved from $c_0$ to $c_1$}
					\State $running\_time\textprime_{c_1} := running\_time_{c_1} + exec_{t_0, c_1}$
					\If{$running\_time\textprime_{c_1} < running\_time_{c_0}$}\label{balance:check_running_time}
						\State $T\textprime_{c_0} \gets T_{c_0} - t_0$
						\State $T\textprime_{c_1} \gets T_{c_1} \cup \{t_0\}$
						\State $T_{b} \gets (T_{b} - \{T_{c_0}, T_{c_1}\}) \cup \{T\textprime_{c_0}, T\textprime_{c_1}\}$
						\State \Goto{balance:select_vm}
					\EndIf
				\EndIf
			\EndFor
			\State \Return $T_{b}$
		\EndFunction
	\end{algorithmic}
\end{algorithm}

Since Algorithm \ref{nearest_plan} and Algorithm \ref{reduce_time_blocks} selects a VM based on the network distance, it is possible that the tasks are not evenly distributed on the VMs.

Hence, the \textit{balancing algorithm} presented in Algorithm \ref{balance} is employed to balance tasks on all VMs and make their running times almost similar. It is quite similar to Algorithm \ref{reduce_time_blocks} which reduces the number of time blocks, but the cloud VMs are selected differently. VMs with most wasted charged time is selected for reducing the number of time blocks, while the VMs with highest running time is selected in the balancing algorithm (Line \ref{balance:select_vm}).

Furthermore, the algorithm has to ensure that moving tasks does not make the receiving VM's execution time higher than a giving one's (Line \ref{balance:check_running_time}). In other words, the balancing process aims to reduce the overall running time by moving tasks from VMs with high execution time to ones with lower execution time. Moreover, it tries to move tasks to nearest VMs possible in order to further minimise overall running time.

\subsection{Finding Execution Plan Based on Given Trade-off}

A complete solution based on the previous algorithms are presented in Algorithm \ref{complete_algorithm}, which finds the plan taking into account a cost versus performance trade-off provided by a user.

\begin{algorithm}
	\caption{Find Plan Algorithm}
	\label{complete_algorithm}
	\begin{algorithmic}[1]
		\Function{FIND\_OPTIMAL\_PLAN}{$T, C, trans, \beta$}
			\State $P \gets \emptyset$
			\State $T_n \gets GET\_NEAREST\_PLAN(T, C, trans)$\label{optimal:nearest_plan}
			\For{$min\_tb$ from 1 to $tb_{T_n}$}\label{optimal:start_finding}
				\State $T_{ce} \gets REDUCE\_TB(trans, T_n, min\_tb)$
				\State $T_b \gets BALANCE(trans, T_{ce})$
				\If{$T_b \notin P$}
					\State $P \gets P \cup \{T_b\}$
				\EndIf
			\EndFor\label{optimal:end_finding}
			\State $max\_tb \gets \max_{T_b \in P}{tb_{T_b}}$
			\State $max\_overall\_exec \gets \max_{T_b \in P}{overall\_exec_{T_b}}$
			\State $T_o = \argmin_{T_b \in P}{(\beta \times \frac{overall\_exec_{T_b}}{max\_overall\_exec}}$\par
	        \hskip\algorithmicindent$+ (1 - \beta) \times \frac{tb_{T_b}}{max\_tb})$\label{optimal:get_plan}

			\State \Return $T_o$
		\EndFunction
	\end{algorithmic}
\end{algorithm}

The inputs of the algorithm are the list of tasks (i.e. their location and data size), cloud VMs, the distances between nodes and the $\beta$ values. First of all, the nearest plan is created (Line \ref{optimal:nearest_plan}). Then, it is used to find the list of plans with difference $min\_tb$ values running from 1 to the $nearest\_plan$'s number of time blocks (Lines \ref{optimal:start_finding} to \ref{optimal:end_finding}). The score value is calculated based on $\beta$ and Equation \ref{score}. Finally, the plan with lowest score is returned as an optimal plan (Line \ref{optimal:get_plan}).

Algorithm \ref{complete_algorithm} can be modified in order to find an execution plan based on given budget constraints, which is converted into the $min\_tb$ value. Hence, instead of using different $min\_tb$ values, the algorithm needs to only one $min\_tb$ value, which is derived from given budget constraint, to find an optimal plan.

\subsection{Optimal Centralised Plan}

\begin{algorithm}
	\caption{Find Optimal Centralised Plan Algorithm}
	\label{centralised_algorithm}
	\begin{algorithmic}[1]
		\Function{FIND\_CENTRALISED\_PLAN}{$T, C, trans$}
			\For{$c \in C$}
				\State $T_c \gets T$
			\EndFor
			\State $T_{centralised} \gets \argmin_{c \in C}{exec_{T_c}}$
			\State \Return $T_{centralised}$
		\EndFunction
	\end{algorithmic}
\end{algorithm}

Algorithm \ref{centralised_algorithm} compares the decentralised and centralised approaches and finds an optimal decentralised execution plan. The idea is to assign the tasks to all VMs and select the VM with lowest execution time.

\section{Experimental Studies}
\label{experimentalstudies}

\subsection{Setup}

To evaluate our approach, we developed a word count application which downloaded text files from globally distributed hosts and counted the number of words in each file without aggregating the result. There were 3290 counting tasks, each of which consisted a text file distributed in one of 47 different nodes on PlanetLab (PL), a global research network which offers over a thousand of globally distributed machines for experimentation \cite{Chun:2003:POT}. We used Amazon Web Service (AWS)\footnote{http://aws.amazon.com/} which offers eight different regions for cloud VMs; thus there are 8 Cloud VMs (instance). Given the list of tasks, the communication cost and the computational cost, the BoDT is divided into smaller subsets, each of which is sent to an application deployed on the AWS instance. Since AWS instance is charged by hour, each $time\_block$ equals to 3600 seconds.

Prior to the experiment, we measured the communication cost (i.e. $trans_{l_l, c}$) and the computational cost (i.e. $comp$) by running a small test in which the applications were deployed in all AWS regions and retrieved and processed data from all PlanetLab nodes.

In order to investigate the trade-off between performance and cost, instead of providing the $\beta$ value and obtaining the corresponding $optimal\_plan$, we executed all possible execution plans, each of which had a different number of time blocks. The $centralised\_plan$ was executed in order to compare the centralised and decentralised approaches.

In order to automate the experiments, we implemented a Scala framework which can find solutions based on a given input, deploy and monitor the execution.

Our framework was able to find 6 different decentralised solutions represented as $plan\_i$ for $3 \geq i \geq 8$, where $i$ represents the number of VMs required, and predicted the number of used time blocks since no instance took more than one hour to finish its execution. Hence, it was predicted that at least 3 time blocks were required to execute the BoDT. We also found a $centralised\_plan$ which was predicted to use 4 time blocks. Each plan was executed three times and the average results are presented.

\subsection{General Results}

\begin{figure}[h!]	\centering
		\includegraphics[width=0.5\textwidth]{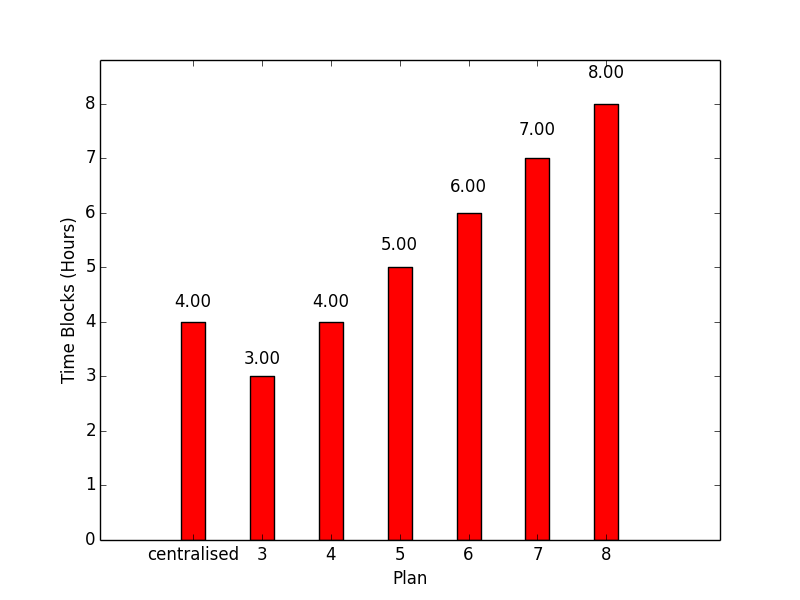}
	\caption{Time blocks for centralised and decentralised plans}
	\label{time_blocks}
\end{figure}

Figure \ref{time_blocks} presents the cost of executing distributed tasks on the cloud with respect to the number of time blocks used by each plan.
It shows that our framework predicted the number of used time block for all plans accurately. The decentralised approach represented by $plan_3$ is cheaper than the centralised one that requires 4 hours. In other words, the benefit of reducing the network distance can outweigh the $deploy\_time$ suffered by each additional VM.

\begin{figure}[h!]
	\centering
		\includegraphics[width=0.5\textwidth]{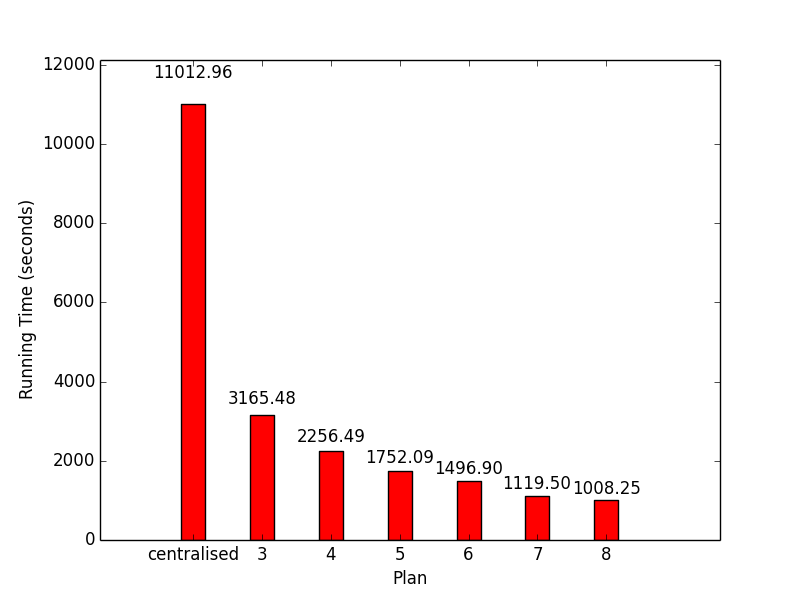}
	\caption{Running time of different plans}
	\label{running_times}
\end{figure}

In order to evaluate the performance of different plans, we measured the overall running times and plotted them in Figure \ref{running_times}. As predicted, using more instances results lower running time. The $centralised\_plan$ has the highest running time, since it does not execute in parallel and most tasks are not assigned to their nearest instances.

Figure \ref{time_blocks} and Figure \ref{running_times} also highlight the trade-off between different decentralised plans; higher the cost with respect to the number of time blocks better performance with respect to execution time is obtained. Moreover, it also reflects the decision that user has to make: whether she favours performance over monetary cost or vice versa.

\subsection{Cost vs Performance Gain}

\begin{figure}[h!]
	\centering
		\includegraphics[width=0.5\textwidth]{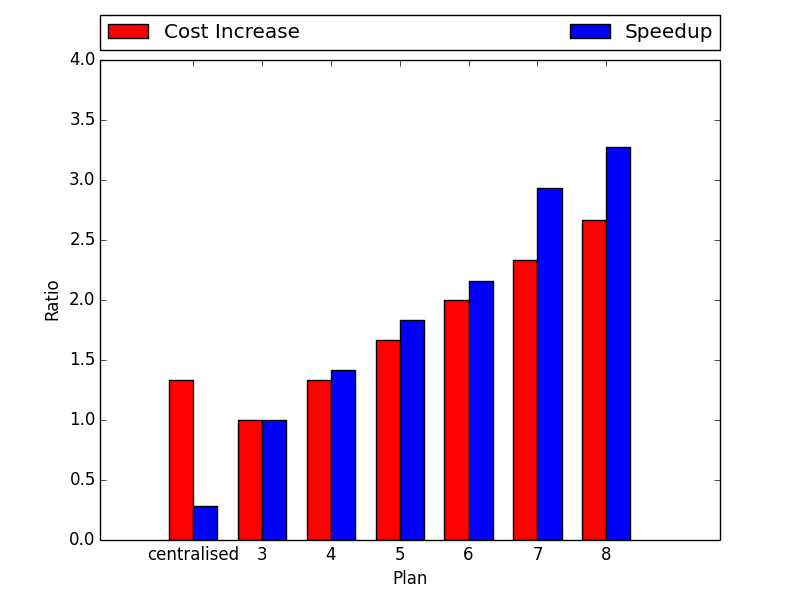}
	\caption{Increase in cost vs speedup for different plans}
	\label{cost_speedup}
\end{figure}

We further investigate the relationship between the increase in cost and performance, for example, whether doubling the cost would provide a twofold performance. For this, $plan\_3$ is selected as the base value, the cost increase is calculated as $\frac{tb_{plan}}{tb_{plan\_3}}$. The performance gain is calculated as the speedup when using $plan_3$ over other plans, i.e $speedup = \frac{running\_time_{plan\_3}}{running\_time_{plan}}$.

Figure \ref{cost_speedup} presents the comparison between the increase in cost and performance gain. As predicted, $centralised_plan$ performs worst as it does not have any speedup in spite of the increase in cost. On the other hand, for the decentralised plans, it is interesting to see that the relationship between cost increment and performance gain is not proportional. In other words, the performance gain is higher than the additional cost.

\begin{figure}[h!]
	\centering
		\includegraphics[width=0.5\textwidth]{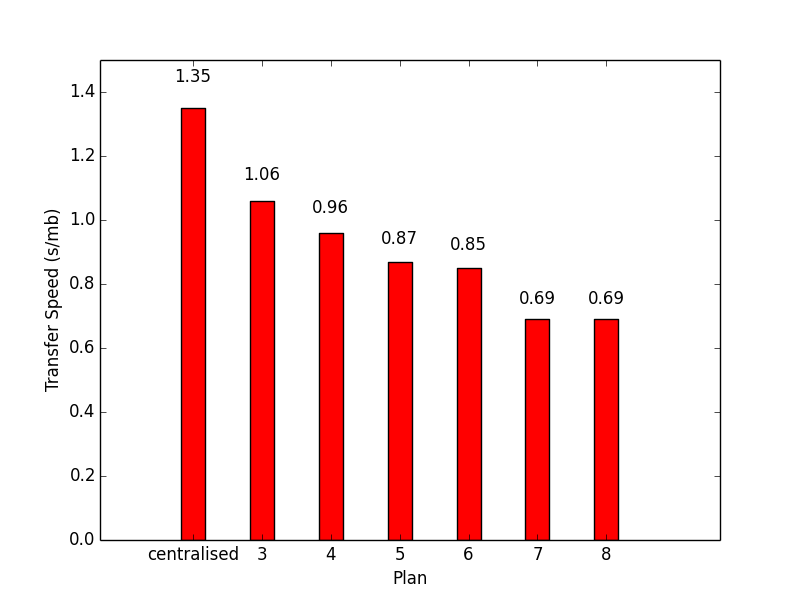}
	\caption{Average data transfer speeds for different plans}
	\label{comm_speed}
\end{figure}

As we used cloud instances of the same type, the computational capacity is identical. However, in BoDT applications, the location of an additional instance also affects the performance gain. While moving tasks from one cloud instance to another, the algorithm always tries to preserve data proximity, i.e. keeping the data transfer time of each task as low as possible. Figure \ref{comm_speed} presents the average data transfer speeds for all execution plans. It shows that expensive plans (with more cloud instances) have lower data transfer speed (high data proximity). So, by employing more instances the overall performance is boosted by not only increasing computational capacity but also reducing the data transfer time.

\subsection{Effect of $\beta$ Value}

\begin{figure}
	\centering
		\includegraphics[width=0.5\textwidth,height=0.25\textheight]{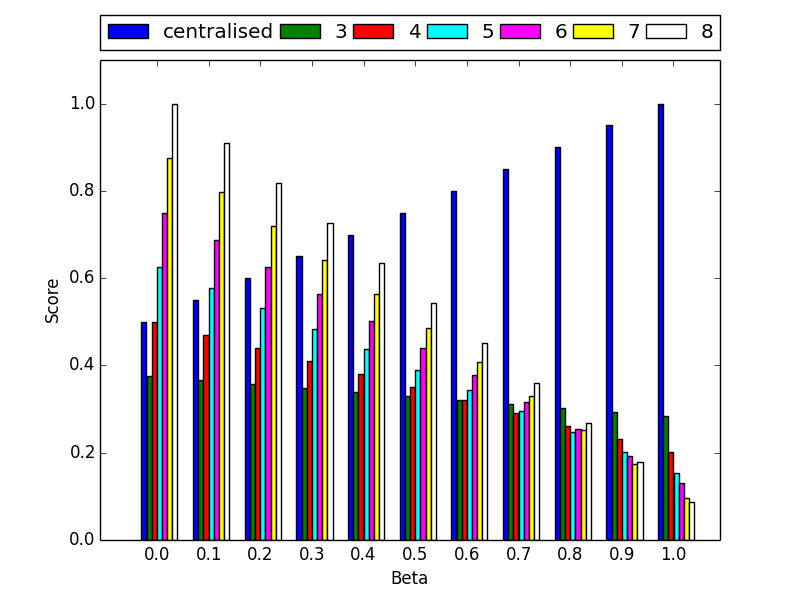}
	\caption{Scores of different $\beta$ values for different plans}
	\label{scores}
\end{figure}

Figure \ref{scores} presents the scores calculated based on Equation \ref{score} with different values of $\beta$ for all plans. It shows how $\beta$, i.e. the importance of performance over cost, affects the scores and the decision of selecting a plan based on user's preference. When $\beta = 0$, (performance is completely ignored in order to minimise cost), $plan\_3$ has the lowest score while $plan\_8$ has the highest score. As $\beta$ is increased the score of  $plan\_8$ is seen to decreases and is the lowest when $\beta = 1.0$. In all cases there is at least one solution which is better than a $centralised\_plan$, i.e. decentralised approaches results in better performance and are cost-saving.

Table \ref{predict_actual} is an evaluation of our approach and its accuracy of prediction of the deployment plan based on the user provided $\beta$ value. The actual plan is the solution with the lowest score. The key result is that the approach makes correct predictions 10 out of 11 times.

The prediction is inaccurate in one case when $\beta = 0.6$. This is due to network instability on the PL nodes resulting in slow data transfer. Hence, VMs that download data from the PL nodes took more time to finish their execution than others. This issue will be investigated further in future work.


\begin{table}
	\centering
	\begin{tabular}{| c | c | c | c |}
		\hline
		$\beta$ & Prediction & Actual & Accurate \\
		\hline
		$0.0$ & $plan\_3$ & $plan\_3$ & Yes \\
		\hline
		$0.1$ & $plan\_3$ & $plan\_3$ & Yes \\
		\hline
		$0.2$ & $plan\_3$ & $plan\_3$ & Yes \\
		\hline
		$0.3$ & $plan\_3$ & $plan\_3$ & Yes \\
		\hline
		$0.4$ & $plan\_3$ & $plan\_3$ & Yes \\
		\hline
		$0.5$ & $plan\_3$ & $plan\_3$ & Yes \\
		\hline
		$0.6$ & $plan\_4$ & $plan\_3$ & No \\
		\hline
		$0.7$ & $plan\_4$ & $plan\_4$ & Yes \\
		\hline
		$0.8$ & $plan\_5$ & $plan\_5$ & Yes \\
		\hline
		$0.9$ & $plan\_7$ & $plan\_7$ & Yes \\
		\hline
		$1.0$ & $plan\_8$ & $plan\_8$ & Yes \\
		\hline
	\end{tabular}
	\caption{Predicted plan vs actual plan for different $\beta$ values}	
	\label{predict_actual}
\end{table}

\section{Related Work}
\label{relatedwork}

Optimising the deployment of an application on the cloud aims to select instance types, number of instances and their locations so that certain criteria such as performance or cost can be satisfied. The work presented in \cite{Tordsson:2012:CBM,Lucas-Simarro:2013} focused on maximising performance while minimising cost and satisfying other constraints.

Moreover, the geographical distribution of services and users are also considered while deploying an application on the cloud in order to satisfy predefined SLA \cite{Zang:2012:ICDCS} or minimise network distance between all nodes \cite{Quia:2013:ACSA}. Similarly, \cite{Kand:CLOUD:2012} and \cite{Zhu:SOSE:2013} considered the relationship between services in order to minimise the communication costs not only from users to services but also between services. In \cite{Luckeneder:2013:CloudCom}, the geographical locations of web services were in into account in order to select a location to deploy workflow orchestration. Finally, in \cite{Ryden:2012:IC2E}, the authors used edge cloud to process large amount of data based on geographical and network distances from the requesting clients, the proposed framework was able to outperform other similar systems.

Besides optimising the deployment of the application, scheduling tasks execution is also important. Data locality and proximity were investigated in \cite{Ranganathan:2002:DCD,Zaharia:2010:DSS}, and it was concluded that the total performance could be increased by reducing the distance between the computation and its required data. In \cite{Oprescu:2010:CloudCom}, the authors used statistical methods to dynamically schedule the execution of tasks on the cloud to satisfy budget constraint and completion time goal.

Stream processing in geographically distributed environment was investigated in \cite{Tudoran:2013:SGS}. The authors proposed a service oriented framework to process and aggregate streaming data from globally located locations.

More recently, there are studies about the trade-off between performance and cost of executing application on the cloud. In \cite{Genez:UCC:2013}, the authors investigated the usage of public cloud, which would increase the cost, to meet the application deadline. Jung el al. proposed a novel workflow scheduler which considered both the execution time and cost of cloud resources \cite{Jung:ICWS:2013}.


Different from existing research which use the trade-off between cost and performance as a constraint to optimise the deployment and/or execution of the application, our goal is to investigate it, i.e. comparing the relationship between the performance gain and monetary cost between different setups. Moreover, the geographical locations of data is also taken into account as they directly affect the overall execution time. Finally, we also compare the centralised and decentralised approaches in order to emphasise the benefit of decentralising of execution on the cloud.

\section{Conclusions}
\label{conclusions}

Cloud computing offers the flexibility of acquiring resources without any prior commitment. Moreover, it provides different types and locations of resource for users to select from in order to maximise an application's performance. The flexibility offered by cloud computing has changed the way the industry as well as the scientific community approach their problems. However, it also introduces new questions, which is how to balance the performance gain and monetary cost of running application on the cloud.

In this paper, we investigated the trade-off between performance gain and monetary cost of executing BoDT on the cloud. Taking the location of tasks into account, we proposed a heuristic algorithm which can find where the tasks should be executed in order to satisfy user's preference of cost versus performance. The trade-off has been clearly presented in our experiment which shows the shifting between different plans when the user's preference changes. The benefit of decentralising the execution instead of running everything in the centralised location is also demonstrated.

In the future, we plan to investigate deploying multiple instances in the same region by taking a variety of cloud VMs that have different performance capabilities and costs into account. Dynamic scheduling will be investigated in order to handle undesired events, for example, network instability during execution.

\section*{Acknowledgments}

This research is supported by the EPSRC grant `Working Together: Constraint Programming and Cloud Computing' (EP/K015745/1), a Royal Society Industry Fellowship `Bringing Science to the Cloud', an EPSRC Impact Acceleration Grant (IAA) and an Amazon Web Services (AWS) Education Research Grant.

\bibliographystyle{ieeetr}
\bibliography{references}

\end{document}